# Hydrogen peroxide forms spontaneously in water (bulk, film, or microdroplet) via reduction of dissolved oxygen at solid-water interface

Muzzamil Ahmad Eatoo[a,b,c,d] and Himanshu Mishra[a,b,c,d]*

[a]Environmental Science and Engineering (EnSE) Program, Biological and Environmental Science and Engineering (BESE) Division,

[b]Water Desalination and Reuse Center (WDRC), King Abdullah University of Science and Technology (KAUST), Thuwal, 23955-6900, Kingdom of Saudi Arabia

[c]Center for Desert Agriculture (CDA), King Abdullah University of Science and Technology (KAUST), Thuwal 23955-6900, Saudi Arabia

[d]Interfacial Lab (iLab), King Abdullah University of Science and Technology (KAUST), Thuwal 23955-6900, Saudi Arabia

*Correspondence: himanshu.mishra@kaust.edu.sa

Zare & co-workers have recently claimed that hydrogen peroxide is spontaneously generated on the air–water interface of sprayed microdroplets, i.e., that $H_2O_2$ forms without an external energy source or co-reactant or catalyst(1). Specifically, they find that the $H_2O_2$(aq) concentration in sprayed microdroplets increases by a factor of 3.5-× (or 2.5-×) as the spray chamber's relative humidity (RH) is changed from 15% to 50% (or from 15% to 95%)(1). Building on these results, they imply causation for the seasonality of viral infections arising from the RH-dependent $H_2O_2$ generation in environmental microdroplets (1, 2). Here, we present an alternative explanation for their observations.

Since 2019 we have been investigating the various claims of spontaneous $H_2O_2$ production at the air–water interface of microdroplets, for example: (i) ~30 μM $H_2O_2$ in sprays(3), (ii) up to 114 μM $H_2O_2$ in condensates at 55% or 70% RH(4), and (iii) up to 180 μM $H_2O_2$ in condensates formed at 50% RH(2). Our investigation, however, revealed that these high levels of $H_2O_2$ were largely formed due to the reaction of water with the ambient ozone gas (5). That is, we found that the role of the air–water interface was limited to accommodating ozone gas, which is minimally soluble in bulk water, and the product of the reaction was $H_2O_2$(aq) that is miscible with water. In support of our arguments, when Zare & co-workers repeated their spray(6) and condensation(7) experiments in an ozone-free environment, they found $H_2O_2$(aq) concentration **dropped drastically** to 0.3–1.5 μM (1, 6). Nevertheless, they concluded that their initial claims about the air–water interface's ability to spontaneously form $H_2O_2$(aq) were correct. To this end, they have suggested a variety of mechanisms, including charge transfer between positively and negatively charged spray microdroplets, as put forth by Colussi(8), to instantaneous ultrahigh electric fields predicted by Head-Gordon & co-workers(9).

Our latest study has unveiled that even this trace-level (0.3–1.5 μM) $H_2O_2$(aq) formation in sprayed or condensed microdroplets does not take place at the air–water interface(10). Instead, it takes place at the numerous water–solid interfaces involved in a typical experiment, such as pipes, tubing, containers, vials, etc., wherein the dissolved oxygen in water ($O_2$(aq)) gets reduced and the solid surface(s) gets oxidized(10). Solids with a higher propensity to be oxidized produce higher amounts of $H_2O_2$(aq), which can be explained via the classical Galvanic series (Fig. 1A). If the $O_2$(aq) concentration is increased, the concentration of $H_2O_2$(aq) follows, which is what Zare & co-workers' observed when the nebulizing gas was changed from $N_2$(g) to $O_2$(g) and the $H_2O_2$(aq) concentration in the sprays increased from 0.3 µM to 1.5 µM (6). Conversely, if O$_2$(aq) is eliminated from the water (by first boiling it and then and bubbling N$_2$(g) for 45 min), then the water–solid interface does not produce any detectable H$_2$O$_2$(aq) (limit of detection: 50 nM) (Fig. 1A inset and Fig. 1B).

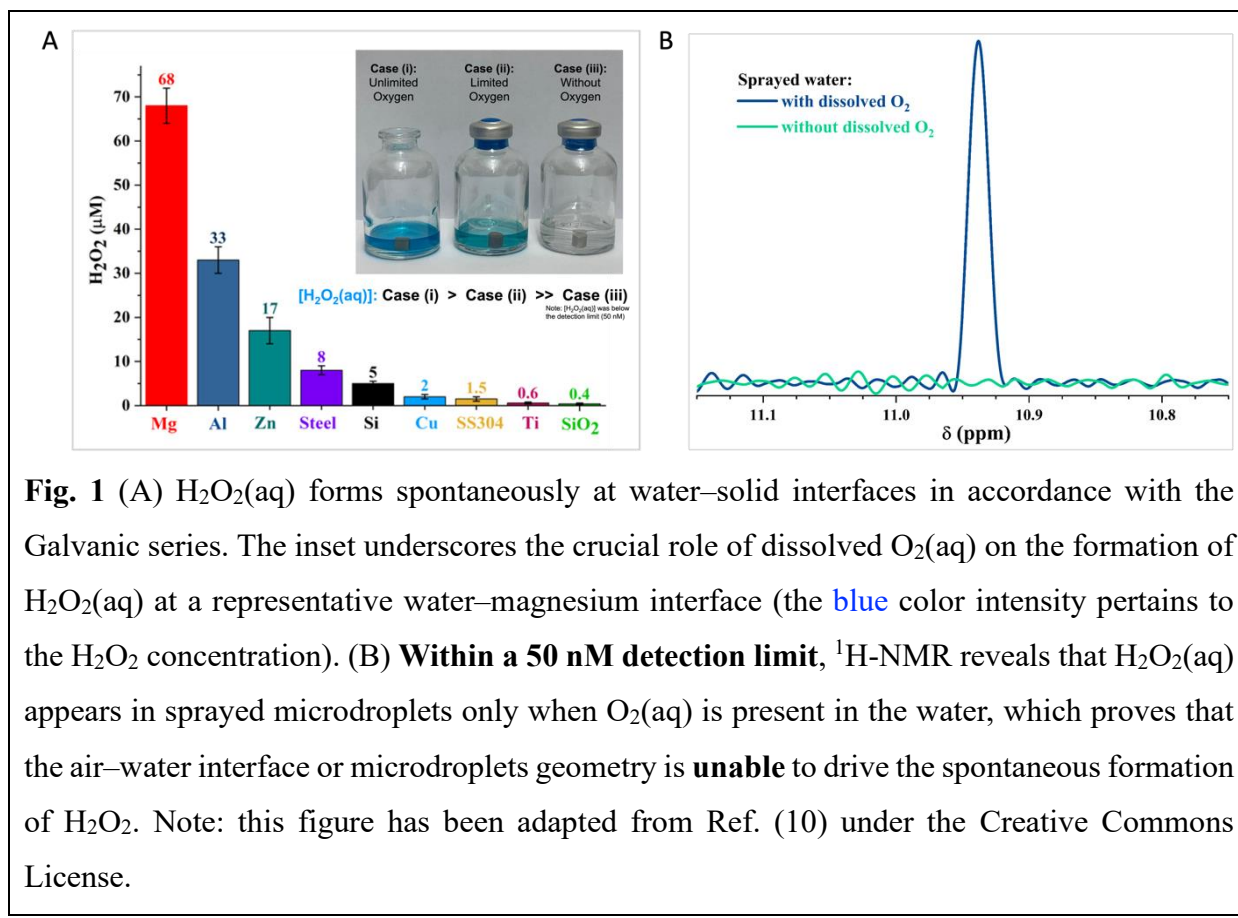

**Fig. 1** (A) H$_2$O$_2$(aq) forms spontaneously at water–solid interfaces in accordance with the Galvanic series. The inset underscores the crucial role of dissolved O$_2$(aq) on the formation of H$_2$O$_2$(aq) at a representative water–magnesium interface (the blue color intensity pertains to the H$_2$O$_2$ concentration). (B) **Within a 50 nM detection limit**, $^1$H-NMR reveals that H$_2$O$_2$(aq) appears in sprayed microdroplets only when O$_2$(aq) is present in the water, which proves that the air–water interface or microdroplets geometry is **unable** to drive the spontaneous formation of H$_2$O$_2$. Note: this figure has been adapted from Ref. (10) under the Creative Commons License.

Given the above, we submit that the latest claims for 0.3–1.5 µM H$_2$O$_2$ formation at the air–water interface and the effects of RH and microdroplet geometry upon this formation are all easily explained as arising instead from redox reactions taking place at the water-solid interface. This, in turn, suggests a recalibration of the environmental and practical implications of this phenomenon, such as for seasonal spread of diseases, and green chemistry.

**Acknowledgements:** HM acknowledges KAUST for funding (Grant# BAS/1/1070-01-01).

**Author contributions:** HM and ME wrote this comment.

**Competing interests:** The authors declare no competing interests.